\input{aipcheck}

\newcommand{\apj}{ApJ}
\newcommand{\apjl}{ApJL}
\newcommand{\mnras}{MNRAS}

\newcommand{\nat}{Nat.}
\newcommand{\jcap}{JCAP}
\newcommand{\physrep}{Physics Reports}

\documentclass[
    ,final            
  ]
  {aipproc}

\usepackage{amssymb}
\usepackage{amsmath}

\layoutstyle{6x9}

\begin{document}

\title{Analytical Modeling of Galaxies at $z\gtrsim6$: Star Formation and Black Hole Growth}


\author{Joseph A. Mu\~{n}oz}{
  address={Department of Physics and Astronomy, University of California Los Angeles, Los Angeles, CA 90095, USA}
}

\begin{abstract}

Galaxies at $z\gtrsim6$ represent an important evolutionary link between the first galaxies and their modern counterparts.  Modeling both the global and internal properties of these recently discovered objects can lead us to understand how they relate to even earlier systems.  I show how the balance of cold inflows and momentum-driven super-winds can explain the evolution of the UV mass-to-light ratio from $z\sim6$--10.  I then describe a model for maintaining hydrostatic equilibrium and marginal Toomre-instability by radiation pressure in dust-free galactic disks.  Applying this framework to $z\sim6$--8 systems, I show how the internal ISM physics can be constrained by X-rays observations with {\it{Chandra}}.

\end{abstract}

\maketitle


\section{Introduction}

Observations with WFC3 have recently discovered large sample of galaxy candidates out to $z\sim10$ using the Lyman-break technique \cite[e.g.,][]{Bouwens06, Bouwens11a, Bouwens11b, Oesch12}.  These systems represent an important evolutionary link between the earliest collections of stars and gas into dark matter halos and the more modern incarnations of the last ten billion years.  Progress in modeling the physics of $z\gtrsim6$ galaxies is intimately related to the state of observations.  Samples with hundreds of such systems has already allowed quite a bit of success in understanding their global and ensemble properties, such as their luminosity functions \cite[e.g.,][]{ML11, Munoz12}.  As an illustration of this, I describes agreement between fits of the observed luminosity function and a simple model of galactic inflow and outflow in predicting the redshift evolution of the UV mass-to-light ratio.  On the other hand, the limited spacial resolution of the observations \cite{Oesch10b} as well as the difficulties inherent in obtaining spectra of these extremely faint objects has made a relative mystery of their internal properties.  To begin to rectify this deficiency on the theoretical side, I present a new model for the maintenance of hydrostatic equilibrium and marginal Toomre-instability in high-redshift disks by radiation pressure that describes their interstellar media, the growth of their central black holes, and the mechanisms that transport angular momentum.  Ultimately, understanding this physics will reveal whether $z\sim6$--$10$ are more similar to their $z\sim2$--$3$ analogs or to their as yet unknown precursors at $z\sim10$--$20$.  

\begin{figure}\label{fig:L10}
 \includegraphics[height=.3\textheight]{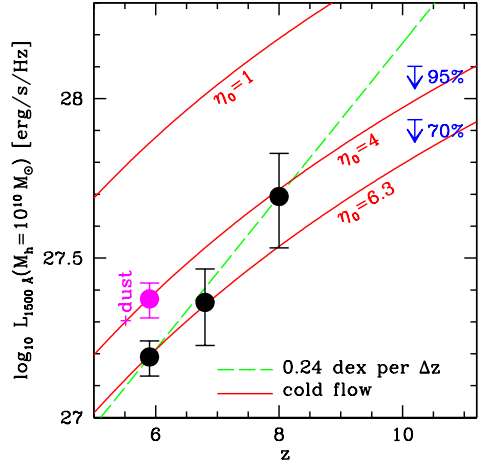}
 \caption{
 The evolution in $L_{10}$, the average UV luminosity of $10^{10}\,{\rm M_{\odot}}$ halos, from $z\sim6$--$10$.  The points and upper limits show fits to the observed luminosity function using the model of \cite{ML11}.  The dashed curve shows a simple extrapolation between the points at $z\sim6$ and 8, while the solid curves show the predictions from Eq. \ref{eq:Luv} for $\eta_0=1$, 4, and 6.  Figure is from \cite{Munoz12}.
 }
\end{figure}

\section{Evolution of the Mass-to-Light Ratio}\label{sec:MtoL}

While the strong luminosity function evolution from $z\sim4$--$10$ can be primarily seen as an expression of the behavior of the dark matter halo mass function \cite{ML11}, it is clear that there exists a residual change in the average mass-to-light ratio.  Figure \ref{fig:L10} shows the increase in mean UV luminosity at fixed halo mass from $z\sim6$--$8$ after the changing halo abundance has been removed \cite{Munoz12}.  However, the trend clearly does not extend out to $z\sim10$ where the paucity of observed galaxies places strong upper limits on the average brightness.  Since describing early galaxies in terms of the inflow and outflow processes that have been explored at lower redshifts is a important way of understanding the evolution of galaxy formation throughout cosmic time, we attempt to calculate the average UV luminosity as a function of halo mass and redshift by balancing cold-flow accretion \cite{Keres05, McBride09} with star formation and momentum driven outflows \cite{Murray05} in equilibrium \cite{Dave12}.  The resulting luminosity is given by \cite{Munoz12}
\begin{equation}\label{eq:Luv}
L_{1500}=\frac{A\,M_{\rm cf}}{1+\eta_{\rm w}},
\end{equation}
where $M_{\rm cf}$ is the average cold-flow accretion rate as a function of halo mass and redshift, and we have assumed a constant conversion factor, $A$, between star formation rate and the luminosity at 1500 {\rm \AA}.  $A\approx8\times10^{27}\,{\rm erg/s/(M_{\odot}/yr)}$ for a Salpeter initial mass function at solar metallicity \cite{Madau98}.  We further parameterize the wind mass-loading factor, $\eta_{\rm w}$, as 
\begin{equation}\label{eq:eta_w}
\eta_{\rm w}=\eta_{\rm 0}\,\frac{100\,{\rm km/s}}{\sigma},
\end{equation}
where $\sigma$ is the halo velocity dispersion as a function of mass and redshift \cite{BL01}.  Figure \ref{fig:L10} shows that this description of the evolving mass-to-light ratio accurately predicts that obtained from fitting the observed luminosity function for a value of $\eta_0\approx4$ when 0.18 dex of dust extinction is applied to the point at $z\sim6$.  The need for these ``super-winds" is consistent with results from numerical simulations at all times \cite{OD08}.  

\begin{figure}\label{fig:Lx}
 \includegraphics[height=.3\textheight,trim=0 260 0 0,clip]{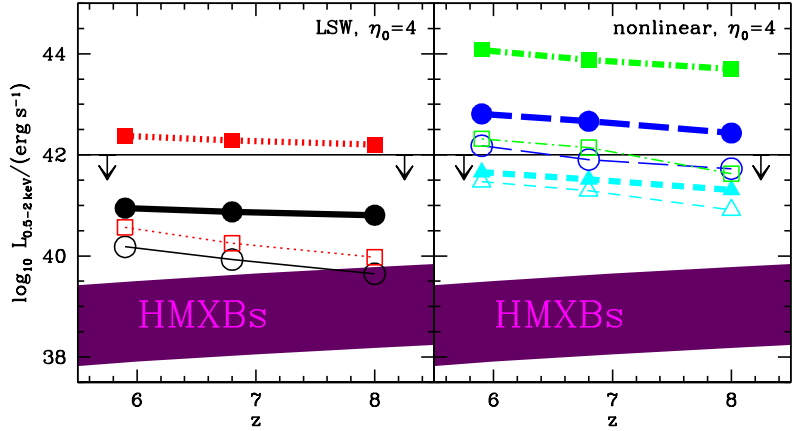}
 \caption{
The average X-ray luminosity from the stacked samples of $z\sim6$, 7, and 8 galaxies described by \cite{Cowie12}.  The horizontal line shows the observed upper limits, while the labeled bands show the expectations from high-mass X-ray binaries.  The left panel shows our model predictions in a linear spiral wave scenario with solid and dotted lines denoting results for $m=0.2$ and $m=1$, respectively.  In the right panel, short-dashed, long-dashed, and dot-dashed lines show results for a nonlinear infall model with $\beta=0.001$, 0.01, and 0.1, respectively.  In both panels, thin lines assume a conservative amount of disk obscuration, while results in thick lines assume no obscuration.  Figure is from \cite{MF12a}.
   }
\end{figure}

\section{Angular Momentum Transport and AGN Fueling}\label{sec:ism}

In the dense interstellar media of $z\sim2$ starbursts, the hot medium of supernova remnants is a negligible component of the pressure equilibrium \cite{Thompson05}.  Rather, radiation pressure from starlight, as well as mechanical pressure from supernova blast waves, generate the turbulence required to balance hydrostatic equilibrium and maintain marginal Toomre-instability (i.e. $Q\sim1$).  This sets the star formation rate as a function of galactic radius independent of the formation micro-physics.  Despite their much lower star formation rate densities, the environments of $z\gtrsim6$ galaxies are also extremely dense.  Thus, in \cite{MF12a}, we adopt this formalism to describe their interiors.  However, because of the inferred lack of dust extinction in these very high-redshift systems, we assume that the radiation pressure is due to photo-ionizations on neutral hydrogen and ignore the effect of IR photon trapping among dust grains.  After accreting onto the galaxy at the cold-flow rate \cite{McBride09}, gas is transported toward the center of the disk.  Along the way, the flow is depleted by star formation and momentum-driven super-winds with $\eta_0=4$.  The remaining gas is accreted onto the central black hole and powers an active galactic nucleus with a radiative efficiency of $\sim10\%$.  The more slowly the gas is transported, the more star formation and outflows diminish the amount available to fuel the black hole.  Using a template spectrum extrapolated from lower redshift, we calculate the X-ray luminosity of the stacked samples of Lyman-break galaxies at $z\sim6$, 7, and 8 observed by \cite{Cowie12} for different phenomenological models of angular momentum transport models in the disk: either for a linear spiral wave (LSW) model where the infall velocity is a constant fraction, $m$, of the local sounds speed or a non-linear, shocked infall model where the infall velocity is a constant fraction $\beta$ of the local circular velocity.  Figure \ref{fig:Lx} compares these results (in both an obscured and un-obscured case) to recent limits by \cite{Cowie12} and to the expected emission from high-mass X-ray binaries.  Observations already rule out a shocked infall scenario in which $\beta=0.1$ and the resulting X-ray emission is un-obscured by the galactic disk.  Deeper X-ray observations and larger sample sizes will soon probe the more reasonable range of parameters space and reveal insights into the inner natures of these early galaxies.

\begin{theacknowledgments} 
It is a pleasure to thank Kazu Omukai and the local organizing committee for a wonderful conference.  I also thank Steve Furlanetto and Avi Loeb for collaboration on the research cited herein.
\end{theacknowledgments}



\bibliographystyle{aipproc}   

\begin{thebibliography}{23}
\expandafter\ifx\csname natexlab\endcsname\relax\def\natexlab#1{#1}\fi
\providecommand{\enquote}[1]{``#1''}
\expandafter\ifx\csname url\endcsname\relax
  \def\url#1{\texttt{#1}}\fi
\expandafter\ifx\csname urlprefix\endcsname\relax\def\urlprefix{URL }\fi
\providecommand{\eprint}[2][]{\url{#2}}

\bibitem[{Bouwens} et~al.(2006)]{Bouwens06}
R.~{Bouwens}, G.~{Illingworth}, J.~{Blakeslee}, and M.~{Franx}, \emph{\apj}
  \textbf{653}, 53--85 (2006).

\bibitem[{Bouwens} et~al.(2011{\natexlab{a}})]{Bouwens11a}
R.~J. {Bouwens}, G.~D. {Illingworth}, I.~{Labbe}, P.~A. {Oesch}, M.~{Trenti},
  C.~M. {Carollo}, P.~G. {van Dokkum}, M.~{Franx}, M.~{Stiavelli},
  V.~{Gonz{\'a}lez}, D.~{Magee}, and L.~{Bradley}, \emph{\nat} \textbf{469},
  504--507 (2011{\natexlab{a}}).

\bibitem[{Bouwens} et~al.(2011{\natexlab{b}})]{Bouwens11b}
R.~J. {Bouwens}, G.~D. {Illingworth}, P.~A. {Oesch}, I.~{Labb{\'e}},
  M.~{Trenti}, P.~{van Dokkum}, M.~{Franx}, M.~{Stiavelli}, C.~M. {Carollo},
  D.~{Magee}, and V.~{Gonzalez}, \emph{\apj} \textbf{737}, 90
  (2011{\natexlab{b}}).

\bibitem[{Oesch} et~al.(2012)]{Oesch12}
P.~A. {Oesch}, R.~J. {Bouwens}, G.~D. {Illingworth}, I.~{Labb{\'e}},
  M.~{Trenti}, V.~{Gonzalez}, C.~M. {Carollo}, M.~{Franx}, P.~G. {van Dokkum},
  and D.~{Magee}, \emph{\apj} \textbf{745}, 110 (2012).

\bibitem[{Mu{\~n}oz} and {Loeb}(2011)]{ML11}
J.~A. {Mu{\~n}oz}, and A.~{Loeb}, \emph{\apj} \textbf{729}, 99 (2011).

\bibitem[{Mu{\~n}oz}(2012)]{Munoz12}
J.~A. {Mu{\~n}oz}, \emph{\jcap} \textbf{4}, 15 (2012).

\bibitem[{Ouchi} et~al.(2005)]{Ouchi05b}
M.~{Ouchi}, T.~{Hamana}, K.~{Shimasaku}, T.~{Yamada}, M.~{Akiyama},
  N.~{Kashikawa}, M.~{Yoshida}, K.~{Aoki}, M.~{Iye}, T.~{Saito}, T.~{Sasaki},
  C.~{Simpson}, and M.~{Yoshida}, \emph{\apjl} \textbf{635}, L117--L120 (2005).

\bibitem[{Mu{\~n}oz} and {Loeb}(2008)]{ML08a}
J.~A. {Mu{\~n}oz}, and A.~{Loeb}, \emph{\mnras} \textbf{385}, 2175 (2008).

\bibitem[{Mu{\~n}oz} et~al.(2010)]{Munoz10}
J.~A. {Mu{\~n}oz}, H.~{Trac}, and A.~{Loeb}, \emph{\mnras} \textbf{405},
  2001--2008 (2010).

\bibitem[{Trenti} et~al.(2012)]{Trenti12}
M.~{Trenti}, L.~D. {Bradley}, M.~{Stiavelli}, J.~M. {Shull}, P.~{Oesch}, R.~J.
  {Bouwens}, J.~A. {Mu{\~n}oz}, E.~{Romano-Diaz}, T.~{Treu}, I.~{Shlosman}, and
  C.~M. {Carollo}, \emph{\apj} \textbf{746}, 55 (2012).

\bibitem[{Finlator} et~al.(2011)]{Finlator11}
K.~{Finlator}, B.~D. {Oppenheimer}, and R.~{Dav{\'e}}, \emph{\mnras}
  \textbf{410}, 1703--1724 (2011).

\bibitem[{Salvaterra} et~al.(2011)]{Salvaterra11}
R.~{Salvaterra}, A.~{Ferrara}, and P.~{Dayal}, \emph{\mnras} \textbf{414},
  847--859 (2011).

\bibitem[{Oesch} et~al.(2010)]{Oesch10b}
P.~A. {Oesch}, R.~J. {Bouwens}, C.~M. {Carollo}, G.~D. {Illingworth},
  M.~{Trenti}, M.~{Stiavelli}, D.~{Magee}, I.~{Labb{\'e}}, and M.~{Franx},
  \emph{\apjl} \textbf{709}, L21--L25 (2010).

\bibitem[{Kere{\v s}} et~al.(2005)]{Keres05}
D.~{Kere{\v s}}, N.~{Katz}, D.~H. {Weinberg}, and R.~{Dav{\'e}}, \emph{\mnras}
  \textbf{363}, 2--28 (2005).

\bibitem[{McBride} et~al.(2009)]{McBride09}
J.~{McBride}, O.~{Fakhouri}, and C.-P. {Ma}, \emph{\mnras} \textbf{398},
  1858--1868 (2009).

\bibitem[{Murray} et~al.(2005)]{Murray05}
N.~{Murray}, E.~{Quataert}, and T.~A. {Thompson}, \emph{\apj} \textbf{618},
  569--585 (2005).

\bibitem[{Dav{\'e}} et~al.(2012)]{Dave12}
R.~{Dav{\'e}}, K.~{Finlator}, and B.~D. {Oppenheimer}, \emph{\mnras}
  \textbf{421}, 98--107 (2012).

\bibitem[{Madau} et~al.(1998)]{Madau98}
P.~{Madau}, L.~{Pozzetti}, and M.~{Dickinson}, \emph{\apj} \textbf{498}, 106--+
  (1998).

\bibitem[{Barkana} and {Loeb}(2001)]{BL01}
R.~{Barkana}, and A.~{Loeb}, \emph{\physrep} \textbf{349}, 125--238 (2001).

\bibitem[{Oppenheimer} and {Dav{\'e}}(2008)]{OD08}
B.~D. {Oppenheimer}, and R.~{Dav{\'e}}, \emph{\mnras} \textbf{387}, 577--600
  (2008).

\bibitem[{Thompson} et~al.(2005)]{Thompson05}
T.~A. {Thompson}, E.~{Quataert}, and N.~{Murray}, \emph{\apj} \textbf{630},
  167--185 (2005).

\bibitem[{Munoz} and {Furlanetto}(2012)]{MF12a}
J.~A. {Munoz}, and S.~R. {Furlanetto}, \emph{arXiv:astro-ph/1201.1300}  (2012).

\bibitem[{Cowie} et~al.(2012)]{Cowie12}
L.~L. {Cowie}, A.~J. {Barger}, and G.~{Hasinger}, \emph{\apj} \textbf{748}, 50
  (2012).

\end{thebibliography}

\end{document}